\documentclass[twocolumn,showpacs,superscriptaddress,preprintnumbers,amsmath,amssymb,prc]{revtex4-1}

\usepackage{graphicx}
\usepackage{dcolumn}
\usepackage{bm}
\usepackage{ifpdf}
\usepackage{epstopdf}
\usepackage{slashed}
\usepackage{amsfonts}
\usepackage{mathrsfs}
\usepackage{amssymb}
\usepackage{multirow}
\usepackage{CJK}
\usepackage{xcolor}
\usepackage{subfigure,dcolumn}
\usepackage[T2A,T1]{fontenc}
\usepackage[english]{babel}
\usepackage{amsmath}
\usepackage{listings}
\usepackage{booktabs}
\usepackage{youngtab}
\usepackage{float}

\begin{document}

\title{Elliptic flow in heavy-ion collisions at intermediate energy: the role of impact parameter, mean field potential, and collision term}

\author{Bo Gao}
\affiliation{School of Science, Huzhou University, Huzhou 313000, China}
\affiliation{School of Physical and Electronic Engineering, Shanxi University, Taiyuan 030006, China}
\author{Yongjia Wang}
\email[Corresponding author, ]{wangyongjia@zjhu.edu.cn}
\affiliation{School of Science, Huzhou University, Huzhou 313000, China}

\author{Zepeng Gao}
\affiliation{Sino-French Institute of Nuclear Engineering and Technology, Sun Yat-sen University, Zhuhai 519082, China}
\affiliation{School of Science, Huzhou University, Huzhou 313000, China}

\author{Qingfeng Li}
\email[Corresponding author, ]{liqf@zjhu.edu.cn}
\affiliation{School of Science, Huzhou University, Huzhou 313000, China}
\affiliation{Institute of Modern Physics, Chinese Academy of Science, Lanzhou 730000, China}

\date{\today}

\begin{abstract}

Within the ultrarelativistic quantum molecular dynamics (UrQMD) model, by reverse tracing nucleons that are finally emitted at mid-rapidity (|$y_0$| < 0.1) in the entire reaction process, the time evolution of elliptic flow ($v_2$) of these traced nucleons produced in Au+Au collisions at beam energy of 0.4 GeV$/$nucleon with different impact parameters ($b$) is studied. The initial value of $v_2$ is positive and increases with $b$, then it decreases as time passes and tends to saturate at a negative value. It is found that nucleon-nucleon collisions always depress the value of $v_2$ (enhance the out-of-plane emission), while the nuclear mean field potential may slightly raise the value of $v_2$ during the expansion stage in peripheral reactions. The related density mostly probed by $v_2$ of nucleons at mid-rapidity is found to be $\sim$ 60\% of the maximum density reached during the collisions.
\end{abstract}
\maketitle

\section{Introduction}
Collective flow which characterizes the collective motion of the produced particles in heavy-ion collisions (HICs) is of great importance for studying the underlying physics \cite{reisdorf1997collective,herrmann1999collective,Danielewicz02,Andronic06,heinz2013collective,Lan2022,Wang2022,Shi2021,Lin2021}. It provides indirect access to the properties of the hot and dense matter created in HICs, therefore it has
been extensively studied both theoretically and experimentally over a broad energy range. In HICs at intermediate energies (with beam energy of several hundreds MeV per nucleon), measured experimental data of collective flow are usually compared with corresponding theoretical calculations performed with transport models, in order to infer physical information, such as the nuclear equation of state (EOS), the in-medium nucleon-nucleon (NN) cross section, the nucleon effective mass, the nuclear symmetry energy, see e.g., Refs.\cite{bleicher2022modelling,li2008recent,reisdorf2012systematics,tsang2012constraints,russotto2016results,huth2022constraining,WOLTER2022103962,li2022accessing,wang2018determination,estee2021probing,morfouace2019constraining,xu2019transport,colonna2020collision}.

The directed flow ($v_1$, also called in-plane flow) and the elliptic flow ($v_2$, also called out-of-plane
flow) are two lower-order components of the collective flow that
have been widely studied in HIC at intermediate energies \cite{reisdorf1997collective,reisdorf2012systematics,WOLTER2022103962}. $v_1$ mainly characterizes the particle collective motion in the reaction plane (which is defined by the impact parameter vector $x$ and the beam direction $z$), while $v_2$ describes the particle collective motion in the direction perpendicular to the reaction plane. Elliptic flow is one of the most important observables in HICs not only at intermediate energies but also at relativistic energies. Elliptic flow in HICs at intermediate energies has attracted considerable attention because it exhibits good sensitivity to the not well constrained nuclear equation of state \cite{Danielewicz02,danielewicz1998disappearance,danielewicz2000determination,wang2020study,cozma2018feasibility}. It is found experimentally that elliptic flow depends on beam energy, impact parameter, particle species, and colliding nuclei \cite{reisdorf2012systematics,abdallah2022light,PhysRevLett.125.262301}. From theoretical point of view, $v_2$ strongly relates to the nuclear mean field potential and nucleon-nucleon (NN) collisions \cite{le2016constraining,le2018origin,Li11,wang2014collective,Xu16,zhang2020progress,wang2020application}. However, a quantitative attribution of these effects to $v_2$ is still not very clear.

In the present work, within the ultrarelativistic quantum molecular dynamics (UrQMD) model, the contributions of each issue to $v_2$ of free nucleons which are finally emitted at mid-rapidity are disentangled by recording the change of momenta of these nucleons during the whole reaction process. The paper is organized as follows. In Sec. II, the method to calculate the individual contributions from the nuclear mean field potential and NN collisions to $v_2$ is presented. We then present
results and discussions in Sec. III and end with a
summary in Sec. IV

\section{methodology}
The elliptic flow parameter $v_2$ is the second-order coefficient in the
Fourier expansion of the azimuthal distribution of emitted particles, $v_2 = \left \langle V_2 \right \rangle=\left \langle \frac{p_x^2-p_y^2}{p_t^2} \right \rangle$. Here, $p_x$ and $p_y$ are the two components of the transverse momentum $p_t=\sqrt{p_x^2+p_y^2}$, and the angular bracket denotes an
average over all considered particles of all events. (Throughout this letter, uppercase $V_2$ represents $\frac{p_x^2-p_y^2}{p_t^2}$ of a nucleon). For a certain species of particles produced in HICs with fixed colliding situation, $v_2$ depends both on the rapidity $y_z$ and the transverse momentum $p_t$. Usually, the scaled unit $y_0\equiv y/y_{1cm}$ (the subscript $1cm$ denotes the incident projectile in the center-of-mass system) is used instead of $y_z$, in the same way as done by the FOPI Collaboration \cite{reisdorf2012systematics}. It is known that,
in HICs with mass-symmetric projectile-target combination, $v_2$ is an even function of $y_0$.
$v_2$ of nucleons at mid-rapidity ($y_0$ $\sim$ 0) is of great importance because these nucleons are expected to be emitted from the most compressed region, it carries information about the hot and dense matter that is formed during HICs. Therefore, $v_2$ of free nucleons finally emitted at mid-rapidity (|$y_0$| < 0.1) are focused on in the following.

For studying HICs at intermediate energies, the following density- and momentum-dependent potential form is frequently used in QMD-like models \cite{hartnack1998modelling,hartnack2012strangeness,zhang2020progress,wang2020application},
\begin{equation}\label{eq2}
U=\alpha \cdot (\frac{\rho}{\rho_0})+\beta \cdot (\frac{\rho}{\rho_0})^{\gamma} + t_{md} \ln^2[1+a_{md}(\textbf{p}_{i}-\textbf{p}_{j})^2]\frac{\rho}{\rho_0}.
\end{equation}
In this work, $\alpha$=-398 MeV, $\beta$=334 MeV, $\gamma$=1.14, $t_{md}$=1.57 MeV, and $a_{md}=500$ $c^{2}$/GeV$^{2}$ are adopted in UrQMD model, which yields a soft and momentum-dependent equation of state with the incompressibility $K_{0}=200~\rm{MeV}$. The symmetry potential derived from the SV-sym34 interaction which yields the nuclear symmetry energy with slope parameter of 81.2 MeV is chosen. Besides, the FU3FP4 parametrization \cite{wang2014collective} for the in-medium nucleon-nucleon cross section and an isospin-dependent minimum spanning tree (isoMST) method \cite{PhysRevC.85.051602} for cluster recognition are used. It is found that with appropriate choices of the above parameters, the recent published
experimental data of HICs at intermediate energies can be reproduced fairly well with the UrQMD model
\cite{wang2018determination,li2022accessing,wang2020application}.

In the framework of the UrQMD model \cite{Bass98,Bleicher1999}, the momenta of each nucleon can be changed either by the force caused by the nuclear mean field potential or by nucleon-nucleon (NN) scattering. As a many-body microscopic transport model, it allows recording momentum variations of each nucleon in the mean field propagation and NN scattering. The variation of $v_2$ caused by a collision can be calculated as
\begin{equation}\label{eq2}
\Delta v_2^{\text{coll}}(t)=\left \langle V_2^{\text{aft.coll}}(t) - V_2^{\text{bef.coll}}(t)\right \rangle .
\end{equation}
The average is taken over all the traced nucleons within the time step $\Delta t$ which is set to 1 fm/c in this work.
The variation of $v_2$ due to the mean field potential can be calculated using the momenta of nucleons after and before mean field propagation,
\begin{equation}\label{eq2}
\Delta v_2^{\text{mf}}(t)=\left \langle V_2^{\text{aft.mf}}(t+\Delta t) - V_2^{\text{bef.mf}}(t)\right \rangle.
\end{equation}
Together, the variation of $v_2$ can be obtained as following
\begin{equation}\label{eq2}
\Delta v_2(t)=\Delta v_2^{\text{mf}}(t) +\Delta v_2^{\text{coll}}(t).
\end{equation}

\section{Results and discussions}
\begin{figure*}
    \centering
    \includegraphics[width=0.9\textwidth]{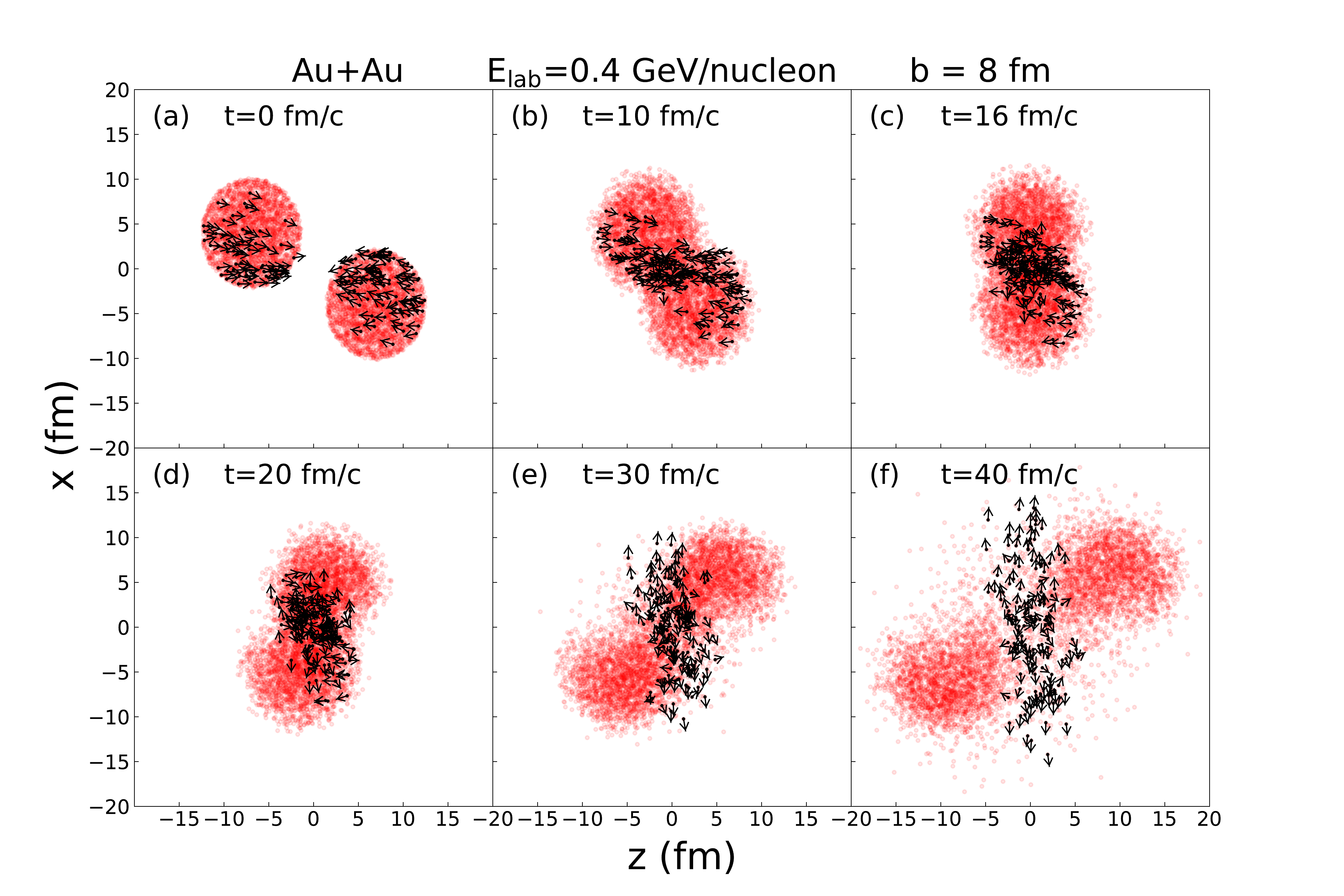}
    \caption{Particle scatter plot in the reaction plane for Au + Au collisions with impact parameter $b$ = 8 fm and beam energy
$E_{\text{lab}}$ = 0.4 GeV$/$nucleon. Solid dot with arrow represents the traced nucleon, while circle denotes other nucleons. The arrow denotes the direction of the momentum vector ($p_x$, $p_z$). Results from 15 random events are displayed. }
    \label{fig:1}
\end{figure*}

It is our goal to quantitatively attribute the effects of initial geometry, mean field potential, and collision term to the formation of elliptic flow in HICs at intermediate energies. In the present work, Au+Au collisions at $E_{\text{lab}}$ = 0.4 GeV$/$nucleon with impact parameters $b$=2, 4, 6, 8, and 10 fm are simulated, momenta of free protons and neutrons finally emitted at mid-rapidity (|$y_0$| < 0.1) are traced during the entire collision process, then the time evolution of $v_2$ of these traced nucleons is investigated. The average number of traced free nucleons are 27.5, 21.2, 14.7, 8.9, and 4.2 for $b$=2, 4, 6, 8, and 10 fm, respectively. More than 50 000 events for each impact parameter are calculated to make the statistical error negligible on the scale of the plots.

Figure \ref{fig:1} displays the location and momentum direction of the traced nucleons in the reaction plane at $t$ = 0, 10, 16, 20, 30, 40 fm$/$c. One sees that the $p_x$ directions of most of the traced nucleons are along the direction
to the coordinate origin, because these nucleons have higher probability to experience a collision, then have higher probability of being emitted in the mid-rapidity at the final state. Collision is the main reason making the initial nucleons (with large |$p_z$| and $y_0 \approx \pm$1) emerge in the mid-rapidity (with almost zero $p_z$ and $y_0 \approx $0) at the final state. In addition, it can be seen that most of the traced nucleons are located around the central region where density is higher than the normal density, while there are still a few of the traced nucleons located away from the central region, see e.g., Fig.\ref{fig:1}(c). This implies that densities probed by $v_2$ of nucleons at mid-rapidity should be smaller than densities around the central region.

\begin{figure}[hbt!]
    \centering
    \includegraphics[width=\linewidth]{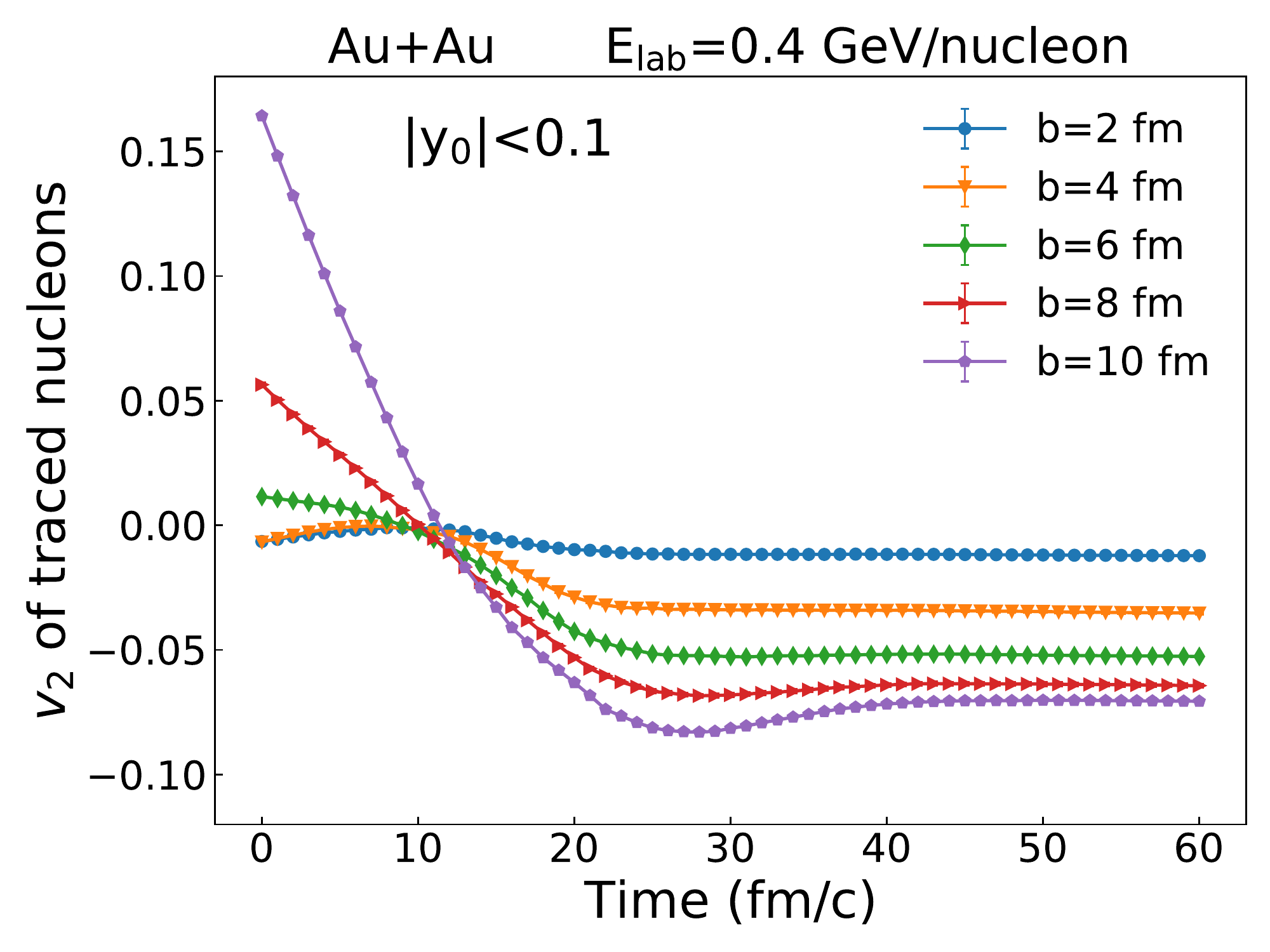}
    \caption{ Time evolution of $v_2$ of the traced nucleons finally emitted at mid-rapidity (|$y_0$| < 0.1) with different impact parameters.}
    \label{fig:2}
\end{figure}

\begin{figure}[hbt!]
    \centering
    \includegraphics[width=0.9\linewidth]{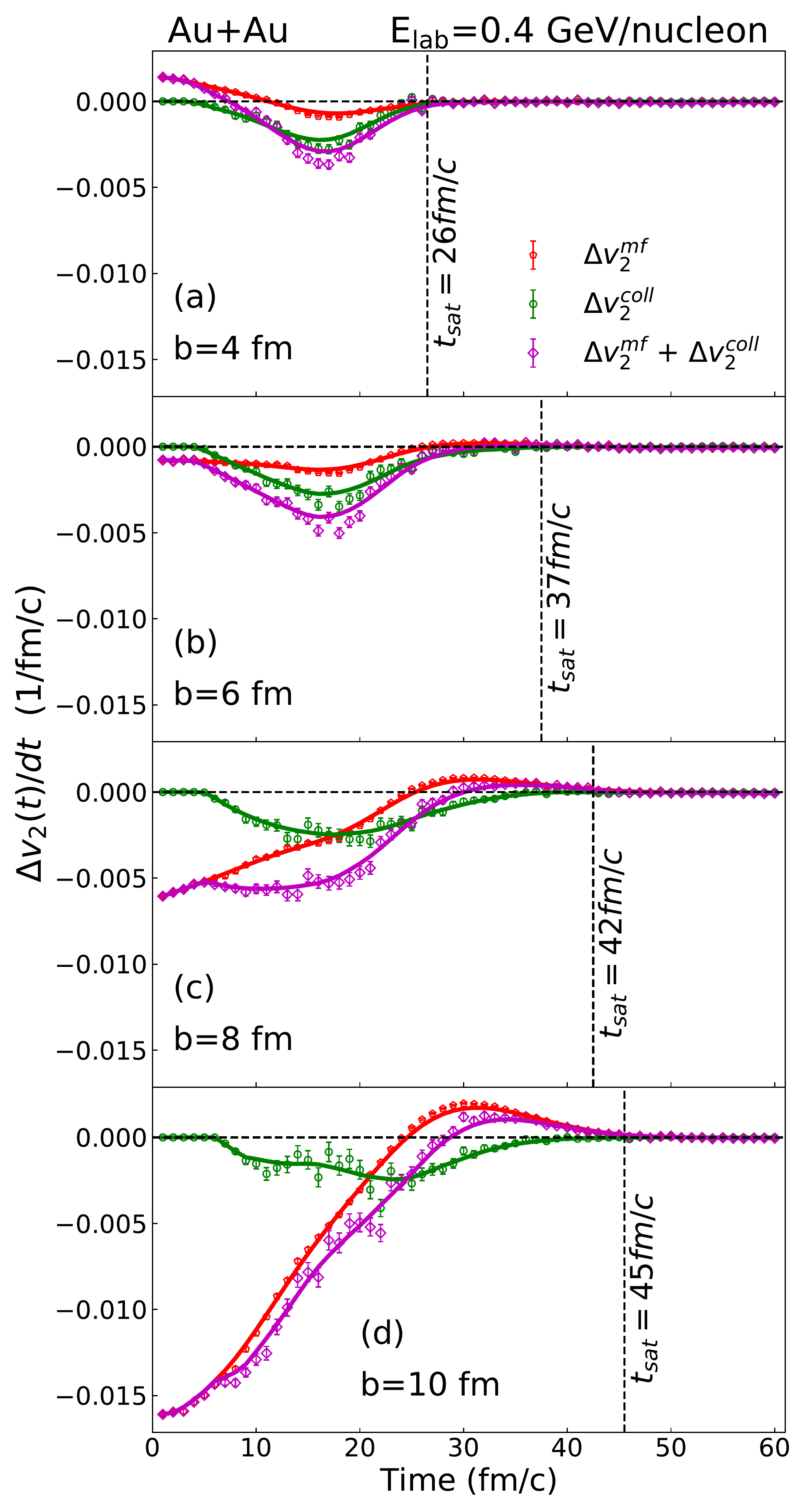}
    \caption{The time evolution of $\Delta v_{2}^{\text {mf}}$(\textit{t}), $\Delta v_{2}^{\text {coll}}$(\textit{t}), and their sum $\Delta v_2$. Lines are smooth curves drawn through the calculated points. The saturated time $t_{\text {sat}}$ for each impact parameter is indicated by italic text along the vertical dashed lines. }
    \label{fig:3}
\end{figure}

Figure~\ref{fig:2} shows the time evolution of $v_2$ of the traced nucleons. At the initial time, $v_2$ is positive (for large $b$) and increases with increasing $b$ because nucleons with larger |$p_x$| and smaller |$p_y$| at $t$=0 fm$/$c tend to have higher probability to experience NN scattering. With a higher $b$, the collision probability decreases accordingly, then it requires nucleons to have much larger |$p_x$| and smaller |$p_y$| at $t$=0 fm/c, so as to enhance the magnitude of $v_2$ at the initial time. For small $b$, all the nucleons have high probability to be collided, thus nucleons appearing in the mid-rapidity at the final state are likely selected randomly, so as to lead to almost a zero $v_2$ at the initial stage. $v_2$ decreases as time passes and approaches to zero at about $t=10 $ fm$/$c, it is mainly caused by the nuclear mean field potential, as only a few collisions
happened at that time, which will be more clearly seen in Fig.\ref{fig:3}. $v_2$ further decreases to negative value and tends to saturate as the reaction
proceeds. In the case of $b$ = 10 fm, the magnitude of $v_2$ first decreases to a minimum value at about $t$ = 25 fm$/$c and then it slightly increases to the saturation value. It is mainly caused by the attraction from target-like and projectile-like fragments in very peripheral collisions.

The contributions of the mean field potential $\Delta v_{2}^{\text {mf}}$(\textit{t}) and NN collisions $\Delta v_{2}^{\text {coll}}$(\textit{t}) to $v_2$ of the traced nucleons are illustrated in Fig.~\ref{fig:3}. Note that $v_2$ in Au+Au collisions with impact parameter $b$=2 fm is too small to get distinct values of $\Delta v_{2}^{\text {mf}}$(\textit{t}) and $\Delta v_{2}^{\text {coll}}$(\textit{t}), thus the result for $b$=2 fm is not presented in the figure.
At the initial time, $\Delta v_{2}^{\text {coll}}$(\textit{t}) is zero because the two nucleons in the first collision must be from target and projectile, it approximately takes 5 fm/c for target and projectile to touch each other. The value of $\Delta v_{2}^{\text {coll}}$(\textit{t}) is always negative which indicates NN collision enhances the signal of elliptic flow (decreases the value of $v_2$ and increases the out-of-plane emission). This is due to the effects of Pauli blocking by the spectator nucleons, because most of the collisions with the momentum of outgoing nucleons along the reaction plane will be blocked. The value of $\Delta v_{2}^{\text {mf}}$(\textit{t}) can be either negative or positive, depending on the impact parameter and time. At the initial time, $\Delta v_{2}^{\text {mf}}$(\textit{t}) is positive (negative) for small (large) $b$ because of the negative (positive) $v_2$, which
can be understood from the fact that the mean field potential tends to attract nucleons until the target and projectile touch with each other. As the collision process gives rise to a region of high density, the mean field potential becomes repulsive and tends to enhance the out-of-plane emission (decrease the value of $v_2$), hence one sees negative $\Delta v_{2}^{\text {mf}}$(\textit{t}). In the cases of $b$=8 and 10 fm, $\Delta v_{2}^{\text {mf}}$(\textit{t}) can become positive during the expansion stage because of the attraction from target-like and projectile-like fragments. The saturation time $t_{\text{sat}}$, which is defined when the absolute value of $\Delta v_{2}$(\textit{t})/$dt$ is smaller than 0.0001 c/fm, increases from 26 fm/c to 45 fm/c when the impact parameter increases from 4 fm to 10 fm.

If one compares $\Delta v_{2}^{\text {mf}}$(\textit{t}) with $\Delta v_{2}^{\text {coll}}$(\textit{t}), it is found that,
the mean field potential plays a dominant role in $v_2$ evolution until $t$$\approx$8 fm$/$c for $b$=4 fm and $\approx$20 fm$/$c for $b$=10 fm. Then the collision effects become dominant and their contributions gradually weaken and eventually disappear as NN scattering ceases. For very peripheral collision, the mean field potential may dominate the evolution of $v_2$ again due to the attraction caused by target-like and projectile-like fragments.

\begin{figure}[hbt!]
    \centering
    \includegraphics[width=\linewidth]{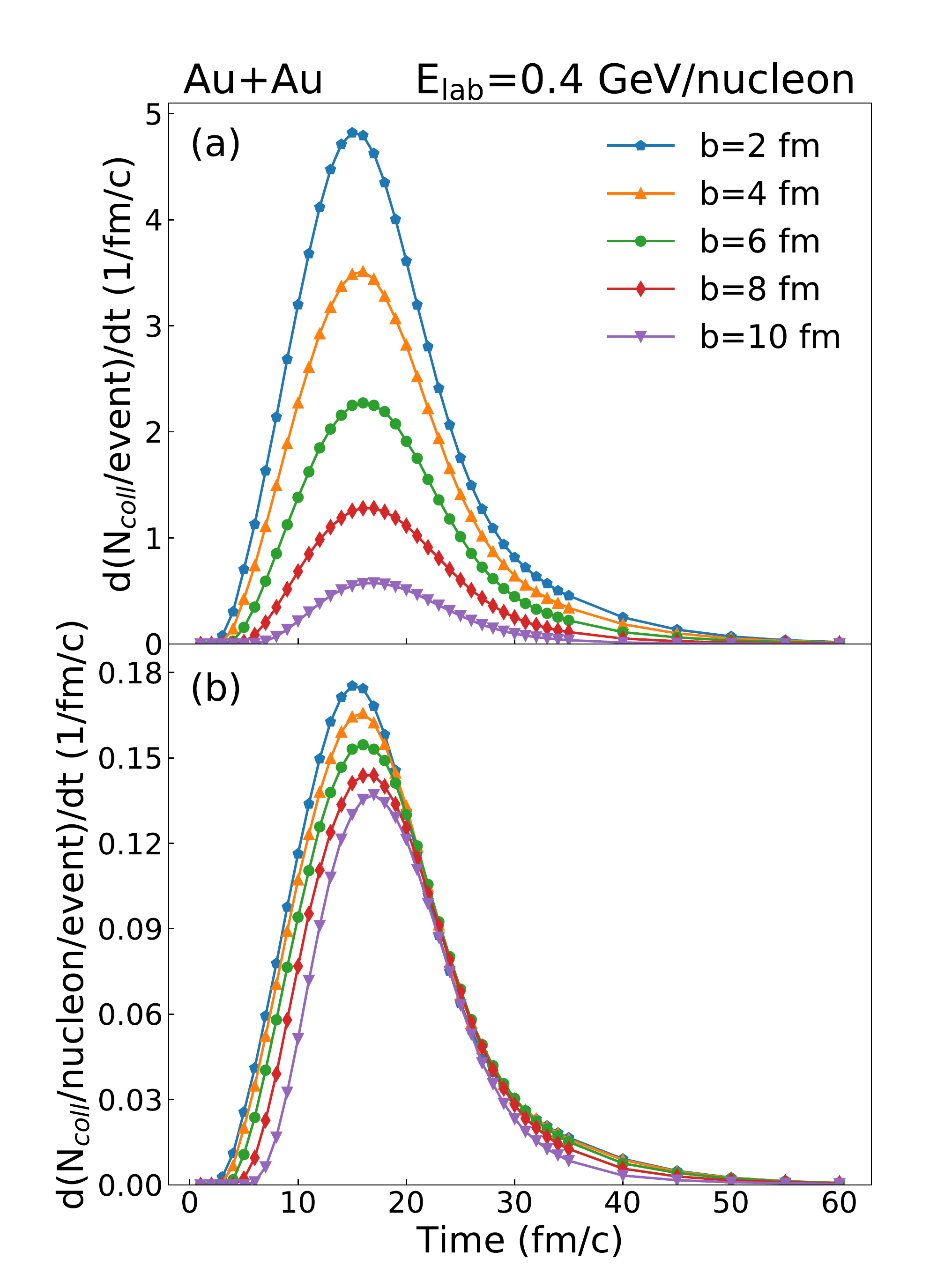}
    \caption{Upper panel: The time evolution of the number of collisions per event experienced by the traced nucleons. Lower panel: The same as the upper panel but for the number of collisions per event per nucleon.}
    \label{fig:4}
\end{figure}

The time evolution of collision rate is important to understand the effect of NN collisions in the evolution of $v_2$. The number of collisions per event and the number of collisions per event per nucleon experienced by the traced nucleons are plotted as a function of time in Fig.~\ref{fig:4} (a) and (b), respectively. It can be seen that, when $b$ is varied from 2 fm to 10 fm, the average collision number per event decreases dramatically, while the average collision number per event per traced nucleon decreases weakly. It is the reason why $\Delta v_{2}^{\text {coll}}$(\textit{t}) in Fig.\ref{fig:3} changes weakly (compared to $\Delta v_{2}^{\text {mf}}$(\textit{t})) with impact parameter. In addition, it is found that there are still a few collisions even after $t_{\text{sat}}$, especially for less peripheral collisions, i.e., $b$=4 fm. These relatively late collisions are weakly blocked by the spectator nucleons, then isotropic scattering is dominant,
thus their contributions to $v_2$ are negligible.

\begin{figure}
    \centering
    \includegraphics[width=\linewidth]{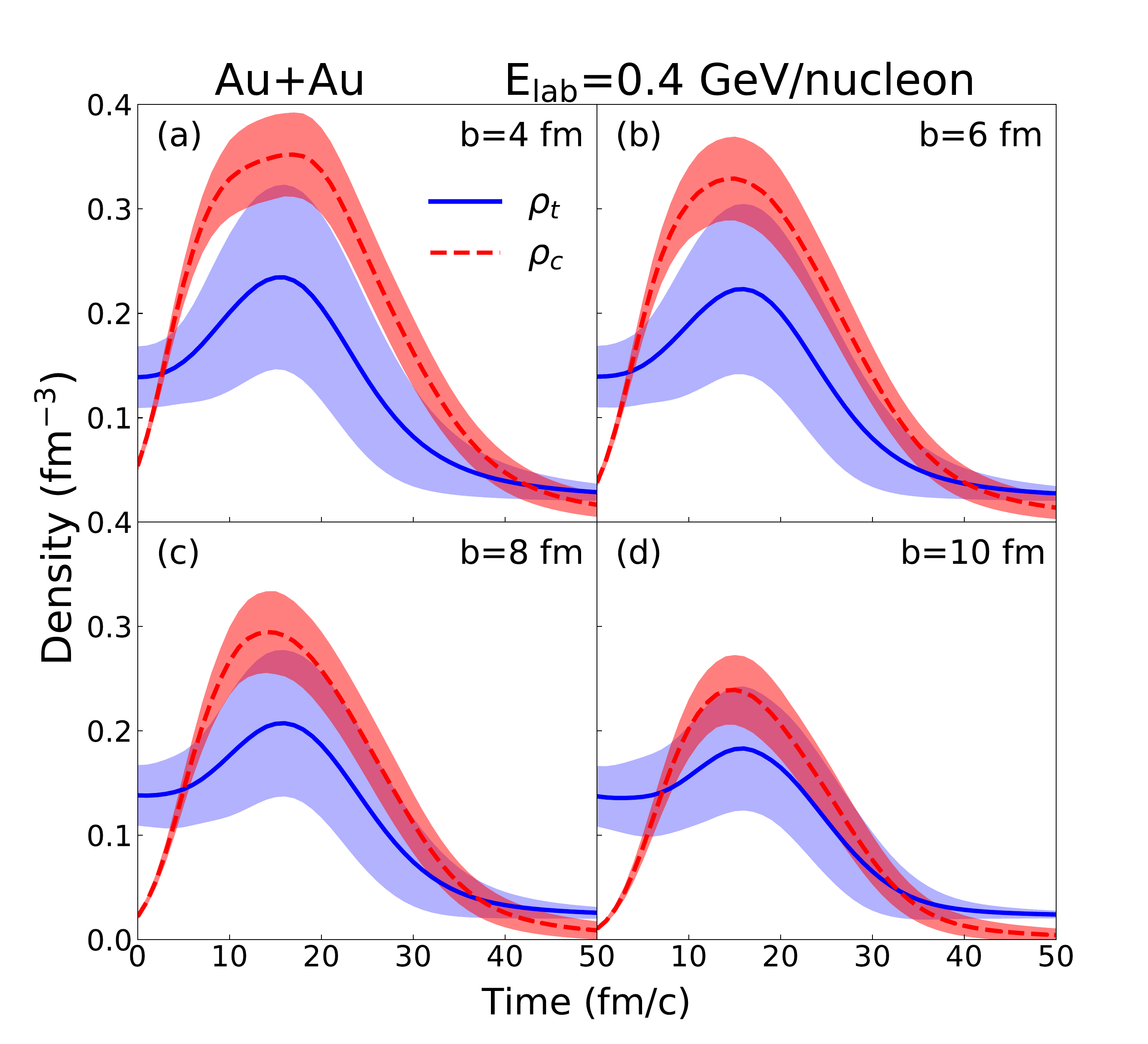}
    \caption{The time evolution of the average density at the central region ($\rho_c$, dashed line) and density experienced by the traced nucleons ($\rho_t$, solid line). The curves and shaded bands denote the mean and the standard deviation, respectively.}
    \label{fig:5}
\end{figure}

It is known that HICs are non-equilibrium and
dynamical processes, the density of the created environment changes with reaction time. It is therefore of interest to know which densities are probed by or mostly related to $v_2$ of nucleons at mid-rapidity. This is an important question especially for investigating the density-dependent nuclear symmetry energy with elliptic flow ratio. To do so, the average density at the central region ($\rho_c$) of Au+Au collisions and density experienced by the traced nucleons ($\rho_t$) are plotted in Fig.\ref{fig:5}. At the initial time, $\rho_c$ is smaller than $\rho_t$ because there are no nucleons located at the coordinate origin at $t$=0 fm$/$c. Further, it can be seen that not only the peak value of $\rho_t$ is smaller than that of $\rho_c$, but also the duration of high density conditions in $\rho_t$ is shorter than that in $\rho_c$. In addition, one finds that the standard deviation of $\rho_t$ is much larger than that of $\rho_c$, because the traced nucleons may come from a variety of spatial regions with different densities, which can be easily understood from Fig.\ref{fig:1}. To quantitatively obtain densities that are mostly related to $v_2$ of nucleons at mid-rapidity, the $\Delta v_2$ weighted average density is calculated as following,
\begin{equation}
\left \langle\rho\right \rangle_{v_2}=\frac{\int_{0}^{t_{sat}} |\Delta v_{2}(t)|\rho_t(t) dt}{\int_{0}^{t_{sat}} |\Delta v_{2}(t)|dt}.
\end{equation}
%


%
The obtained $\left \langle\rho\right \rangle_{v_2}$ are 0.206 and 0.145 fm$^{-3}$ for $b$=4 and 10 fm, respectively. These values are consistent with that obtained in Ref.\cite{le2016constraining}, where the force (due to the nuclear mean field potential) weighted density is obtained to be 0.187 fm$^{-3}$ for the same system and beam energy. Moreover, in Ref.\cite{russotto2016results}, by thoroughly analyzing the elliptic flow difference resulting from switching the density-dependent nuclear symmetry energy at certain density regions, the maximum sensitivity of the neutron-proton elliptic flow ratio lies in the (1.4$\sim$1.5)$\rho_0$ region, this value is close to our result even though a different method and transport model are applied. In Ref.\cite{liu2021insights}, the pion production rate weighted and the force acting on $\Delta$ resonance weighted densities for pion observable are obtained to be 0.272 and 0.224 fm$^{-3}$ for the same system and beam energy but for more central collisions, respectively. These calculations together indicate that the most related density probed by $v_2$ of nucleons at mid-rapidity from Au+Au collisions at $E_{\text{lab}}$ = 0.4 GeV$/$nucleon is around (0.9$\sim$1.3)$\rho_0$ (depending on the impact parameter).  This value is only about 60\% of the maximum density created during the reaction and is smaller than that probed by the pion observable.

\section{Summary}
To summarize, by reverse tracing nucleons finally emitted at mid-rapidity (|$y_0$| < 0.1) in the entire reaction process, the contributions of the nuclear mean field potential and NN collisions to the elliptic flow evolution in Au+Au collisions at beam energy of 0.4 GeV$/$nucleon with different impact parameters are studied.
 At the initial time, $v_2$ of the traced nucleons are positive and increase with increasing $b$, because of the initial Fermi momentum. $v_2$ decreases as time passes and approaches zero at about $t $=10 fm$/$c, it is mainly caused by the nuclear mean field potential. As the reaction
proceeds, the contributions of the NN collisions to $v_2$ become more and more important, $v_2$ further decreases to negative values and tends to saturate. It is found that NN collisions always depress the value of $v_2$, while the mean field potential may slightly raise the value of $v_2$ during the expansion stage in peripheral reactions. The most related density probed by $v_2$ of nucleons at mid-rapidity in Au+Au collisions at 0.4 GeV$/$nucleon lies in (0.9$\sim$1.3)$\rho_0$ and relates to the impact parameter. Its value is found to be $\sim$ 60\% of the maximum density reached during the collisions.
\section{Acknowledgement}
We thank Wolfgang Trautmann for a careful reading of the
manuscript and valuable communications. We acknowledge fruitful discussions with Yingxun Zhang and the TMEP group.
The authors are grateful to the C3S2 computing center in Huzhou University for calculation support. The work is supported in part by the National Natural Science Foundation of China (Nos. U2032145, 11875125, and 12147219), the National Key Research and Development Program of China under Grant No. 2020YFE0202002.
\bibliography{apssamp}
\bibliographystyle{elsarticle-num}

\end{document}